\documentclass[conference,letterpaper]{IEEEtran}

\usepackage[utf8]{inputenc} 
\usepackage[T1]{fontenc}
\usepackage{url}
\usepackage{ifthen}
\usepackage{cite}
\usepackage{graphicx}
\usepackage{textcomp}
\usepackage{xcolor}
\usepackage[utf8]{inputenc} % use UTF8 encoding
\usepackage{kotex} % use KoTeX package for Korean
\usepackage{physics} % 브라켓 및 수식을 위해 사용
\usepackage{amsthm}
\usepackage{algorithm}
\usepackage{algpseudocode}

\theoremstyle{definition}
\newtheorem*{level}{Level}
\newtheorem*{3D}{3-D Potential Graph}
\newtheorem*{VE}{Voltage Estimation}

\theoremstyle{remark}
\newtheorem{df}{Definition}
\begin{document}
\title{Kirchhoff's Circuit Law Applications to Graph Simplification in Search Problems}
%\title{Graph Simplification for Path-finding Problems: A Grafting Electric Potential Analysis Approach}
%\title{Graph Simplification using Electric Potential Theory}
%\title{From Graph Analysis using Electric Potential to Graph Simplification Algorithm using Kirchhoff’s Circuit Law}
% \title{From Graph Analysis using Electric Potential to Graph Simplification via Kirchhoff’s Circuit Law}

% %%% Single author, or several authors with same affiliation:
% \author{%
%   \IEEEauthorblockN{Stefan M.~Moser}
%   \IEEEauthorblockA{ETH Zürich\\
%                     ISI (D-ITET)\\
%                     CH-8092 Zürich, Switzerland\\
%                     Email: moser@isi.ee.ethz.ch}
% }

%%% Several authors with up to three affiliations:
%\author{%
%  \IEEEauthorblockN{Jaeho Choi}
%  \IEEEauthorblockA{Chung-Ang University\\
%                    Seoul, Korea\\
%                    Email: jaehochoi2019@gmail.com}
%  \and
%  \IEEEauthorblockN{Joongheon Kim}
%  \IEEEauthorblockA{Korea University\\
%                    Seoul, Korea\\
%                    Email: joongheon@korea.ac.kr}
%}

\author{\IEEEauthorblockN{Jaeho Choi}
\IEEEauthorblockA{\textit{School of Computer Science and Engineering} \\
\textit{Chung-Ang University}\\
Seoul, Republic of Korea \\
jaehochoi2019@gmail.com}
\and
\IEEEauthorblockN{Joongheon Kim}
\IEEEauthorblockA{\textit{School of Electrical Engineering} \\
\textit{Korea University}\\
Seoul, Republic of Korea \\
joongheon@korea.ac.kr}
}

\maketitle

\begin{abstract}
This paper proposes a new analysis of graph using the concept of electric potential, and also proposes a graph simplification method based on this analysis.
% 이 논문은, 전위의 개념을 사용한 그래프 새로운 해석론을 제안하며, 이 해석론을 바탕으로한 그래프 간소화 기법을 제안한다.
Suppose that each node in the weighted-graph has its respective potential value.
% 먼저 엣지 cost 그래프의 노드별로 각각의 포텐셜를 갖는다고 가정하며,
Furthermore, suppose that the start and terminal nodes in graphs have maximum and zero potentials, respectively. 
% 시작 노드가 가장 높은 포텐셜, 끝 노드가 0의 포텐셜을 갖는다고 가정한다.
When we let the level of each node be defined as the minimum number of edges/hops from the start node to the node, the proper potential of each level can be estimated based on geometric proportionality relationship.
Based on the estimated potential for each level, we can re-design the graph for path-finding problems to be the electrical circuits, thus Kirchhoff's Circuit Law can be directed applicable for simplifying the graph for path-finding problems.
\end{abstract}

\section{Introduction}
The graph as a research topic is still held important positions in various fields, such as mathematics, theoretical computer science~\cite{wang2017graph, molkaraie2013partition,prakash2018coded,host2018coding,anis2018sampling,aggarwal2011achieving}, networks, machine learning, and quantum computing~\cite{duncan2019graph,8849710,choi2019tutorial}.
% 연구 주제로서의 그래프는 수학, 이론 컴퓨터 과학, 네트워크, 머신러닝, 양자 컴퓨팅까지 많은 분야에서 아직까지 중요한 자리를 차지하고 있다.
In particular, many researchers in various fields are working for graph simplification because human beings are faced with complex graphs problems on many issues.
% 특히, 인간들은 많은 이슈에서 복잡한 그래프 문제에 직면하고 있기 때문에, 다양한 분야의 많은 연구자들이 그래프 단순화를 목적으로 연구하고 있다.
Recently, various research topics related to graph simplification have been proposed in many areas.
% 최근, 그래프 단순화와 관련된 다양한 연구 주제가 많은 분야에서 제안되고 있다.
The summary of related research areas about graph simplification is as follows, i.e., data mining~\cite{ruan2011distance}, decentralized composite optimization~\cite{wang2019graph}, network model simplification~\cite{dias2018graph,jongsma2015model,zhonghua2016research,yaw2019graph,dias2017hierarchical}, biological network analysis~\cite{ko2017graph}, scheduling~\cite{sou2014nonserial}, and clustering coefficient~\cite{jung2017sigcon}.
% 그래프 간소화에 대한 최근의 연구 분야를 요약하면 다음과 같다. 데이터 마이닝 분야, 분산 복합 최적화 문제 분야, 네트워크 간소화 분야, 생물 네트워크 분석 분야, 스케줄링 분야, 군집계수 관련 분야.
As such, interest in graph simplification dramatically increases.
% 이렇듯, 그래프 간소화에 대한 관심은 다양한 분야에서 계속되고 있다.
Therefore, this paper proposes a novel graph simplification algorithm.
% 우리 또한, 그래프 간소화에 대한 이러한 지속적인 관심에 동참하여, 새로운 접근방식을 가진 그래프 단순화 알고리즘 한가지를 제안한다. 

The graph simplification algorithm, which is introduced in this paper is based on graph analysis using electric potential.
% 본 논문에서 소개된 그래프 간소화 알고리즘은 전위를 이용한 그래프 분석에 기초한다.
% 우리의 간소화 알고리즘은 경로 탐색 문제를 염두에 두고 그래프에 접근하므로, 알고리즘의 기초가 되는 분석 방법도 항상 시작 노드와 단자 노드를 고려한다.
Our graph analysis method uses the concept of node level, this means the minimum number of edges from the start node to the node.
% 우리의 그래프 분석법은 노드 레벨이라는 개념을 사용하는데, 이는 시작노드로 부터 해당 노드까지의 최소 엣지 개수를 의미한다.
By using the concept of levels and potentials, we create various transformed graphs and present a methodology for analyzing them with a geometric approach.
% 우리는 레벨과 전위를 기반으로 다양한 변형 그래프를 생성하고 기하학적 접근 방식으로 이를 분석하는 방법론을 제시한다.
The ultimate purpose of the methodology that we introduce is to estimate the entire voltage of the graph.
% 우리가 소개하는 방법론의 최종적인 가장 큰 목적은 그래프의 전압 추정이다.
By estimating the entire voltage of the graph and mapping the edge cost of the graph to the resistance, we can estimate the values of each current flowing through each edge using Kirchhoff's Circuit Law.
% 그래프의 전압을 추정하고 그래프의 엣지 코스트를 저항에 대응시켜, 키르히호프 법칙을 이용하여 그래프 내에 흐르는 전류들의 값을 추정해 낼 수 있다.
Then, we can remove the edges, which have no current flow.
% 그런 다음, 우리는 전류가 흐르지 않는 엣지를 제거할 수 있다.
This is an overview of the graph simplification algorithm.
% 이것이 이 논문에서 자세히 설명될 그래프 간소화 알고리즘의 개요이다.

\section{Background}
In this section, we briefly describe two background concepts, i.e., electric potential and Kirchhoff's Circuit Law.
% 이 섹션에서는 우리의 메인 아이디어의 바탕이 되는 두 가지 개념들에 대해 간단히 설명한다.
\subsection{Electric Potential}

%%%% 제거부분
%\subsubsection{Electric Field}
%The electric field is defined as follows:
% 전기장은 다음과 같이 정의된다.
%\begin{equation}\label{eq1}
%    E=\frac{1}{4\pi\epsilon_0}\int\frac{1}{\nu^2}\hat{\nu}dq,
%\end{equation}
%where $\nu=$
%%%% 제거부분

% 전기장은 회전(curl)이 항상 0인 특별한 벡터 함수이다.
The electric field $\vectorbold{E}$ is a special vector function whose curl is always zero~\cite{griffiths2005introduction, serway2018physics} where the $\vectorbold{E}$ can represent as the vector sum of each electric field produced by each charge as:
% 또한 전기장은 다음과 같이 각 전하가 만드는 세부 전기장의 벡터합이다.
\begin{equation}\label{eq1}
    \vectorbold{E}=\vectorbold{E}_1+\vectorbold{E}_2+\cdots+\vectorbold{E}_N,
\end{equation}
where $N$ can be a positive integer.
% 여기서 N은 양의 정수가 될 수 있다.
Thus, the electric field $\vectorbold{E}$ has following properties:
% 따라서 전기장은 다음과 같은 성질을 갖는다.
%\begin{equation}\label{eq2}
\begin{multline}\label{eq2}
    \curl{\vectorbold{E}}=\curl{(\vectorbold{E}_1+\vectorbold{E}_2+\cdots+\vectorbold{E}_N)}\\
    =(\curl\vectorbold{E}_1)+(\curl\vectorbold{E}_2)+\cdots+(\curl\vectorbold{E}_N)=0.
\end{multline}
%\end{equation}
According to \textit{Stokes' theorem}~\cite{spivak2018calculus}, 
% 스토크스의 정리에 따르면, 다음의 식은 식2에 의해 충족된다.
\begin{equation}\label{eq3}
    \oint\vectorbold{E}\vdot d\vectorbold{l}=0
\end{equation}
is satisfied by~\eqref{eq2}. 
By~\eqref{eq3}, following scalar function is defined:
% 식3에 의해, 다음과 같은 스칼라 함수를 정의할 수 있다.
\begin{equation}\label{eq4}
    V(r)\equiv-\int_o^r\vectorbold{E}\vdot d\vectorbold{l},
\end{equation}
where $o$ is a standard reference point, and $r$ is a target point.
% 여기서 o는 기준점이며, r은 타겟점이다.
%%%%%%%%%%%%%%%%%%%%%%%%%%%%%%%%%%%%%%%%%%%%%%%%%%%%%%%%%%%%%%%%%%%%%%%%%%%%%%%%%%%%%%%%%%%%%%%%%%%%%%%% 
%적분 구간에 들어가는 부분이라 r을 그냥 스칼라 표기로 다 바꾸어도 무관할 것 같아서, r에 대한 벡터 표기를 전부 스칼라 표기로 바꾸었습니다.
%%%%%%%%%%%%%%%%%%%%%%%%%%%%%%%%%%%%%%%%%%%%%%%%%%%%%%%%%%%%%%%%%%%%%%%%%%%%%%%%%%%%%%%%%%%%%%%%%%%%%%%%
The scalar function defined in~\eqref{eq4} is called electric potential, and the differential form of~\eqref{eq4} is as follows:
% 식4에서 정의된 스칼라 함수를 전위라고 하며, 식4의 미분꼴은 다음과 같다.
\begin{equation}\label{eq5}
    \vectorbold{E}=-\gradient V.
\end{equation}

In particular, in circuit theory, the difference of electric potential with these characteristics is called voltage.
% 다음과 같은 특징을 갖는 전위의 차이를, 회로 이론에서는 특별히 전압이라 한다.

\subsection{Kirchhoff's Circuit Law}
Kirchhoff's Circuit Law consists of Kirchhoff's Current Law (KCL) and Kirchhoff's Voltage Law (KVL)~\cite{oldham2008doctrine}.
% 키르히호프의 법칙은 전류 법칙과 전압 법칙으로 나뉜다.
%This subsection briefly describes these two sub-laws. % 이 서브섹션에서는 두 가지 개념에 대해 간단히 설명한다.
\begin{itemize}
    \item 
\textit{Kirchhoff's Current Law}: KCL, also called Kirchhoff's first law, states that the sum of incoming currents at a junction is equal to the sum of outgoing currents at the junction.
% 키르히호프의 첫번째 법칙인 이 법칙은 회로의 접합점으로 들어오는 전류의 합과 나가는 전류의 합이 같다는 법칙이다. 
If we define the sign of incoming currents at the junction as positive and the sign of outgoing currents at the junction as negative, this law can be represented that the sum of the currents at each junction is zero as:
% 접합점으로 들어오는 전류의 부호를 +, 접합점에서 나가는 전류의 부호를 -로 두었을 때, 각 접합점에서의 전류의 합이 0 이라는 규칙으로 다음과 같이 간단히 표기할 수 있다. 
\begin{equation}\label{eq6}
    \sum_{n=1}^x \nolimits I_n=0,
\end{equation}
where $I_n$ is each current $n$ and $x$ is the number of incoming or outgoing currents.
% 여기서 I_N은 각 전류를 나타내며, x는 접함점으로 들어오거나 나가는 모든 전류의 개수이다.
In other words, KCL is the same as the law of conservation of charge.
% 즉, KCL은 일종의 전하량 보존 법칙이다.
\newline
    \item
\textit{Kirchhoff's Voltage Law}: KVL, also called Kirchhoff's second law, states that the sum of voltages of the closed-circuit loop is zero.
% 키르히호프의 두번째 법칙인 이 법칙은 폐회로 루프를 이루는 전위차의 총 합이 0 이라는 법칙이다.
If we define the sign of voltage similar to KCL's currents case, this law can be as:
% 전압에 KCL의 전류의 경우처럼 부호성을 부여할 경우, 이 법칙은 다음과 같이 간단하게 표현될 수 있다.
\begin{equation}\label{eq7}
    \sum_{n=1}^y \nolimits V_n=0,
\end{equation}
where $V_n$ is each voltage $n$ and $y$ is the number of voltages measured in the closed-circuit loop.
% 여기서 V_N은 각각의 전압강하를 나타내며, y는 폐회로 루프 속의 측정된 전압강하의 개수이다.
In other words, KVL is the same as the law of energy conservation in a complete closed-circuit.
% 즉, KVL은 폐회로에서의 일종의 에너지 보존 법칙이다.
\end{itemize}

\section{Graph Analysis using Electric Potential}
%\section{Problem Definition and Proposed Algorithm}
%This section discusses about our problem and then explains the proposed algorithm which aims at the conversion of the problem to equivalent easier form.
% 이 섹션에서는 우리가 풀고자 하는 대상 문제에 대해 설명하고, 대상 문제를보다 보다 쉬운 문제로 바꿔줄 우리의 아이디어를 설명합니다.
This section describes the graph analysis method, which combines the concept of each electric potential to each node, before describing the graph simplification algorithm.
% 이 섹션에서는 그래프 간략화 알고리즘을 설명하기에 앞서, 각 노드에 전위의 개념을 도입한 그래프 분석법에 대해 설명한다.
\subsection{Motivation}
%예비 \subsection{Notations and Problem Definition}
Suppose that there is a graph $G(v,e)$ in two-dimensional space where $v$ and $e$ represent nodes and edges, respectively.
% 2차원 평면에 주어진 그래프 G(v,e)가 있다고 하자. 여기서 v는 노드, e는 엣지를 나타낸다.
We have to search for a logically dependable path from the start to the terminal on $G$.
% 우리는 이 그래프 G의 시작 노드에서 터미널 노드까지의 논리적으로 신뢰할 수 있는 경로를 탐색해야 한다.
We already know a variety of simple methods for searching paths, e.g., greedy algorithms~\cite{cormen2009introduction,wilt2010comparison}.
% 우리는 그리디 탐색 등과 같은 다양한 경로 탐색을 위한 알고리즘을 알고 있지만, 
However, if the original graph $G$ is too complex to find the logically dependable path in a simple way, it is essential to simplify the graph before searching for a logically dependable path, e.g., the shortest path, the minimum cost path, and so forth.
% 따라서 신뢰할 수 있는 경로(최단 경로, 최소 코스트 경로) 찾기에 앞서 그래프를 간소화는 필수적이다.
Obviously, prior to simplifying the graph, we need to analyze the graph with various perspectives.
% 당연하게도, 그래프 간소화에 앞서 우리는 다양한 시각으로 그래프를 분석할 필요가 있다.
As one of the new perspectives, we introduce the method of graph analysis using electric potential in the next section.
% 하나의 새로운 시각으로, 다음 섹션에서 우리는 전위를 이용한 그래프 분석법을 소개한다.

\subsection{Graph Analysis using Electric Potential}
We can interpret the graph as an electric circuit by regarding the electric resistances, junctions, and the voltage from the power supply as edges, nodes, and electric potential difference between the start and the terminal.
% 우리는 저항을 엣지에, junction을 노드에, 그리고 전원으로 부터 걸리는 전위차를 시작노드와 끝노드의 전위차로 대응시켜 그래프 마치 전기회로처럼 해석할 수 있다.
If there is a voltage $\varepsilon$ applied to the graph, then the start has an electric potential $\varepsilon$ and the terminal has an electric potential $0$.
% 그래프 전체에 $\varepsilon$의 전압이 걸려 있다고 할 때, 시작노드는 $\varepsilon$의 전위를 갖고, 터미널노드는 0의 전위를 갖으며,
Then, eventually, the rests of the nodes (i.e., non-start and non-terminal nodes) have electric potentials between $\varepsilon$ and $0$.
% 나머지 노드들은 $\varepsilon$와 0 사이 값의 전위를 갖게 된다.

Each junction in the electric circuit has its own electric potential, and no current flows through the resistance between the junctions with the same electric potential.
% 회로에서의 각 junction은 각각의 전위를 갖고, 전위가 같은 junction 사이의 저항에는 어떠한 전류도 흐르지 않는다.
In other words, if there are edges directly connected between nodes with the same electric potential in the graph, no current will flow through the edges.
% 다시 말하면, 그래프 내의 같은 전위를 갖는 노드끼리 직접 연결된 엣지가 있을 경우, 해당 엣지에는 전류가 흐르지 않을 것이다.
Thus, we may remove those edges in the graph.
% 따라서 우리는 해당 엣지를 그래프에서 제거할 수 있다.
By removing the edges between equipotential nodes, the complex graph can be simplified, and the simplified graph has an advantage in the path-finding problem.
% 이렇게, 등전위 노드 사이의 엣지의 제거를 통해, 복잡한 그래프는 간소화될 것이고, 간소화된 그래프는 경로 찾기 문제에서 이점을 갖게 될 것이다.

To find equipotential nodes, the estimation of the electric potential of each node in the graph is required at first.
% 등전위를 갖는 노드를 찾기 위해서는, 먼저 그래프 전체의 노드들이 갖는 각각의 전위에 대한 추정이 필요하다.
The electric potential of each node can be estimated by finding each current value at each edge, and this value can be obtained from KCL and KVL.
% 각 노드들이 갖는 전위는 각 엣지들에 흐르는 전류를 구함으로써 추정할 수 있는데, 이는 키르히호프의 회로 법칙을 통해 구할 수 있다.
However, in order to use KCL and KVL, it is necessary to determine the voltage of the entire graph or the value of the currents diverging from the start.
% 그러나 키르히호프의 회로 법칙을 사용하기 위해서는, 전체 그래프가 갖는 전압의 결정 혹은 시작 노드로 부터 갈라지는 전류들의 값에 대한 결정 등이 필요하다.
For these decisions, we introduce some concepts and analysis methods.
% 이러한 결정들을 위해 몇가지 개념과 분석론을 도입한다.

\begin{figure}[t]
    \centering
    \includegraphics[width=.8\linewidth]{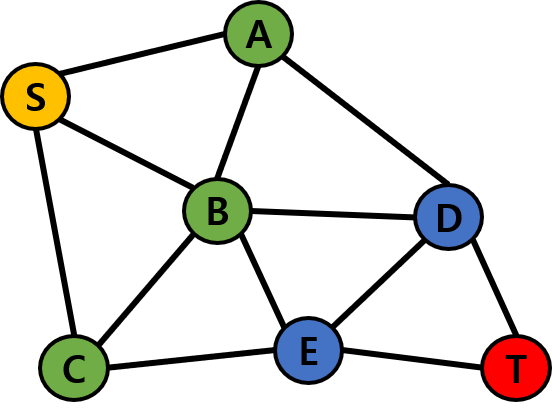}
    \caption{Sample graph $G(7,12)$. The yellow node $S$ represents the start node and the red node $T$ represents the terminal node.
    % 샘플 그래프 G. 노랑노드는 시작점을, 빨간점은 터미널노드를 나타낸다.
    \newline}
    \label{fig1}
    \centering
    \includegraphics[width=.8\linewidth]{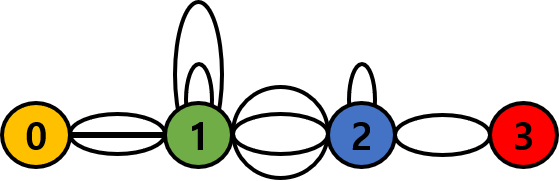}
    \caption{Level unit transformed graph from $G(7,12)$ in Fig.~\ref{fig1}.}
    %그림1의 G 그래프의 레벨 단위 변환 그래프.
    \label{fig2}
\end{figure}

%\textbf{Level.}
\begin{level}
The minimum number of edges between the start node to the target node. 
% 시작점에서 노드까지의 최소 엣지수
\end{level}
In Fig.~\ref{fig1}, the color represents the level of each node.
% 그림1의 예시에서 노드의 색깔은 각 노드의 레벨을 나타낸다.
Level $0$, $1$, $2$, and $3$ correspond to yellow, green, blue, and red, respectively.
% 시작 노드(레벨0 노드)는 노란색, 레벨1 노드는 초록색, 레벨2 노드는 파란색, 터미널노드 (레벨3 노드)는 빨간색이다.
At the graph in Fig.~\ref{fig1}, if we group the nodes at the same level, the graph transforms to level unit graph in Fig.~\ref{fig2}.
% 그림1의 그래프에서, 같은 레벨의 노드끼리 묶으면 그림2의 레벨 단위 그래프로 변환된다.
The number of edges in the transformed graph is equal to the number of edges in the original graph.
% 변환된 그래프의 엣지 개수는 변환전의 그래프와 같다.
The level unit transformed graph can be used in one process of the maximum voltage estimation in the graph simplification.
% 레벨 단위 변환 그래프는 그래프 간소화의 최대 전압 추정의 한가지 과정에서 사용될 수 있다.

\begin{figure}[t]
    \centering
    \includegraphics[width=\linewidth]{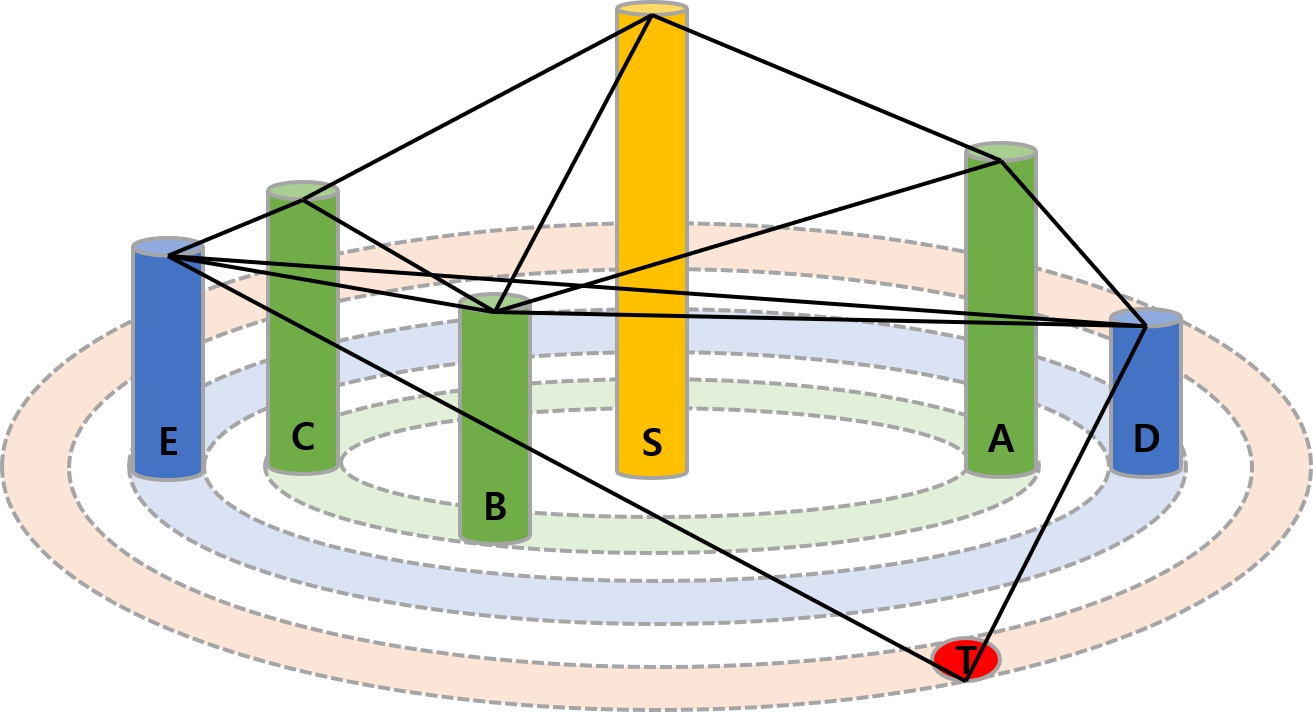}
    \caption{$3$-D potential graph. The original graph in Fig.~\ref{fig1} can be transformed into this graph form. The height of each node column corresponds to each node's electric potential. Each level has a circular orbit, and each node column can be located only at the corresponding level orbit.
    \newline}
    % 3d 전위 그래프. 그림1의 원본 그래프는 다음과 같은 형태로 표현가능하다. 그림1의 각 노드의 전위는 해당 그림의 노드기둥의 높이로 표현된다. 레벨별로 각각의 원 궤도를 갖고, 각 노드 기둥은 해당되는 레벨의 원 궤도에만 위치할 수 있다.
    \label{fig3}
    \centering
    \includegraphics[width=.55\linewidth]{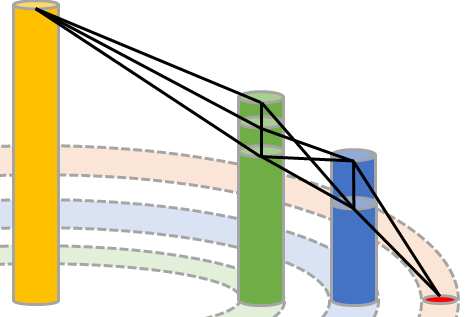}
    \caption{Straight-line alignment by the position moving of each node column at each level orbit in Fig.~\ref{fig3}. The three level $1$ node columns are overlapped and the two level $2$ node columns are overlapped.}
    % 그림 3의 각 레벨 궤도에서 각 노드기둥의 위치 이동에 의한 직선 정렬. 레벨1에는 세 개의 노드기둥 A,B,C가, 레벨2에는 두 개의 노드기둥 D,E가 겹쳐 있다.
    \label{fig4}
\end{figure}

%\textbf{3-D Potential Graph.}
\begin{3D}
The $3$-dimensional graph via mapping nodes' electric potential value to the height value.
% 각 노드의 전위값을 높이로 매핑한 3차원 그래프.
\end{3D}
Each node in a two-dimensional graph is simply represented by a circle or a point, but in a $3$-D potential graph in Fig.~\ref{fig3}, it can be represented by a node column.
% 2차원의 그래프의 각 노드는 원 혹은 점으로 단순하게 표현되지만, 3차원 전위 그래프에서는 이를 노드기둥으로 표시한다.
The representation of the node column's height indicating the electric potential can show visibly the presence of equipotential nodes.
% 전위를 나타내는 노드 기둥의 높이 표현은, 등전위 노드가 존재할 경우 이를 가시적으로 보여준다.
This graph is meaningful in itself but can be interpreted by transforming it as shown in Fig.~\ref{fig4}.
% 이 그래프는 그 자체로도 그래프 해석적 의미가 있지만, 그림4와 같이 변형시켜 해석할 수 있다.
This transformation, aligning all the node columns in a straight line, has several analytical advantages.
% 이러한 변형을 통해, 모든 노드 기둥을 일직선으로 정렬하면, 몇 가지 해석적 이점이 있다.
In Fig.~\ref{fig4}, we can connect the highest point in the start node column where the edges first extend, the lowest point in the terminal node column where the edges finally gather, and the floor point of the start node column.
% 그림 4의 일직선 정렬 변형 그래프에서, 우리는 엣지가 처음 뻗어나가는 시작점, 엣지가 마지막으로 모이는 끝점 그리고 시작 노드 기둥의 바닥지점을 연결할 수 있다.
These three points make a right triangle.
% 이 세점은 하나의 직각 삼각형을 이룬다.
Based on this right triangle, we can add some definitions to estimate the length of each side of the right triangle.
% 이 직각삼각형을 바탕으로, 우리는 직각삼각형의 각 변의 길이를 추정하기 위한 몇 가지 가정을 추가할 수 있다.
And this allows us to determine the voltage of the entire graph.
% 이를 통해 전체 그래프의 전압을 정할 수 있다.

\begin{figure}[t]
    \centering
    \includegraphics[width=\linewidth]{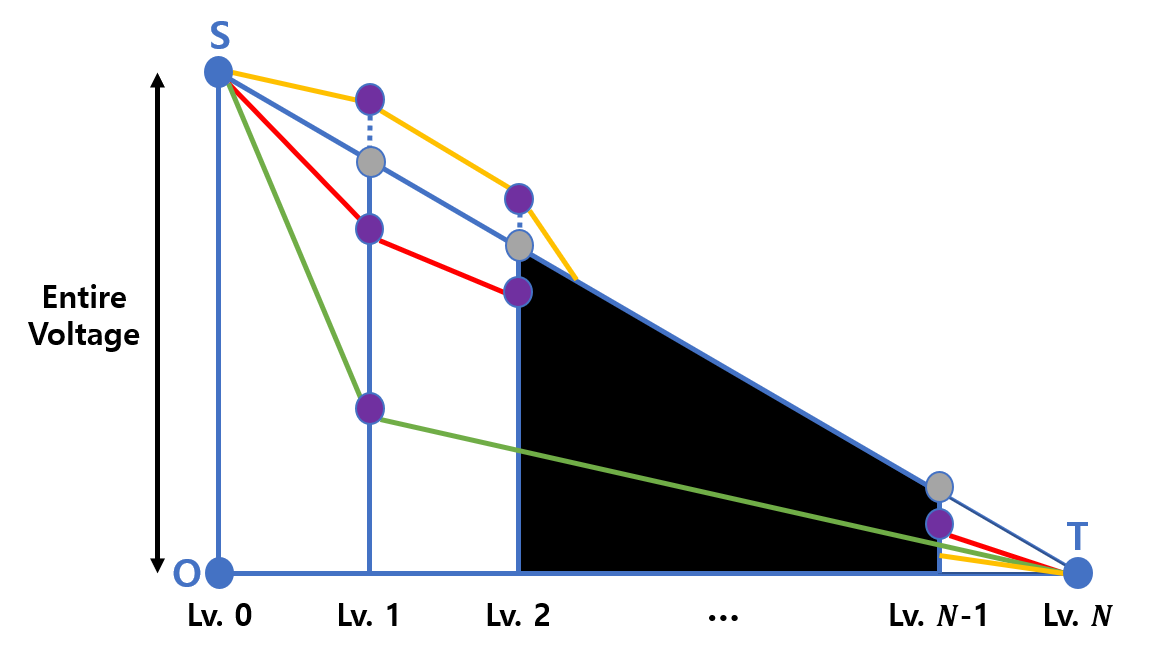}
    \caption{
    Geometric analysis of the $N$-level graph from Fig.~\ref{fig4}.
    % 그림4로 부터 비롯된 기하학적 분석.
    The light blue points $S$, $T$, and $O$ represent the highest point in the start node column, the lowest point in the terminal node column, and the floor point of the start node column, respectively.
    % 연파랑색 S,T,O는 각각 the highest point in the start node column, the lowest point in the terminal node column, and the floor point of the start node column을 나타낸다.
    The purple points represent the highest points of each node columns, and the gray points represent the ideal points at each level.
    % 보라색 점들은 각 노드기둥들의 가장 높은점을 나타내며, 회색점들은 각 레벨에서의 ideal point를 나타낸다.
    The orange, red, and green lines represent examples of different paths, and the black area represents black-box sections between level $2$ and level $N-1$.
    % 주황색,빨간색,초록색 선은 서로 다른 경로의 예시를 나타내며, 검은색 부분은 레벨2~N-1 사이의 블랙박스 부분을 나타낸다.
    }
    \label{fig5}
\end{figure}

\begin{VE}
The entire voltage of the $N$-level graph can be estimated as a positive value less than the sum of the minimum costs between the level $k-1$ and level $k$, where $N$ and $k$ are positive integers, $1\leq k\leq N$.
% 그래프에 걸리는 전체 전압은, 레벨과 레벨 구간 사이의 최소 코스트들의 합보다 작은 양의 값으로 추정 가능.
\end{VE}
Before the description, suppose that there is an $N$-level graph, where $N$ is a positive integer.
% 설명에 앞서, 레벨N의 그래프가 있다고 가정하자.
Transform this graph in the form shown in Fig.~\ref{fig4}, and then connect the highest point in the start node column, the lowest point in the terminal node column, and the floor point of the start node column to make a right triangle.
% 그림 4의 형태로 해당 그래프를 변형하고, the highest point in the start node column, where the edges first extend, the lowest point in the terminal node column, where the edges finally gather, and the floor point of the start node column 세 점을 이어 직각 삼각형을 만든다.  
The right triangle can be created as shown in Fig.~\ref{fig5}.
% 직각삼각형은 그림5와 같이 만들어질 것이다.
The entire voltage of the graph is equal to the electric potential of the level $0$ node, which is equal to the length of $\overline{SO}$.
% 그래프 전체의 전압은 레벨 0 노드의 전위 (최대 전위) 와 같고, 이는 그림 5의 변 SO의 길이와 같다.
The straight line of each level represents each plane perpendicular to the plane where $\triangle{SOT}$ is located.
% 또한 각 레벨의 직선은 삼각형SOT가 위치한 평면과 수직을 이루는 각각의 평면을 나타낸다.
$\overline{ST}$ represents the shortest straight distance passing through the ideal nodes, that maybe existed or not.
% 변ST는 각 레벨의 ideal node를 지나는 최단 직선 거리를 나탄낸다. 
$\overline{ST}$ is shorter than any path, including the orange, red and green lines shown in Fig.~\ref{fig5}.
% 이 직선거리는 그림5에 나타난 주황,빨강,초록 선의 경로들을 포함한 어떤 경로들보다 짧다.
From this, we can make two heuristic definitions.
% 이로 부터, 우리는 두 가지 휴리스틱한 가정을 할 수 있다.

\begin{df}\label{as1}
Let $\overline{ST}$ be the sum of the minimum costs between adjacent levels.
% 변 ST를 그래프의 구간별 최소 코스트들의 합으로 간주하자.
\end{df}
If we expressed the the edge costs between level $k-1$ and level $k$ as $C_{(k-1),(k)}$, then the length of $\overline{ST}$ is as follows:
% 각 레벨 k-1과 레벨 k 사이의 엣지 코스트를 $C_{(k-1)to(k)}$라 표현하면, ST의 길이는 다음과 같다:
\begin{equation}\label{eq8}
\begin{split}
    \overline{ST}= & \min C_{(0),(1)}+\min C_{(1),(2)} + \\
    & \cdots+\min C_{(N-1),(N)}=\sum_{k=1}^N \nolimits \min C_{(k-1),(k)},
\end{split}
\end{equation}
where $min$ means the minimum value.
% 여기서 $min$은 최소값을 나타낸다.
Since $\overline{ST}$ is the length of the longest side of $\triangle{SOT}$, the entire voltage corresponding to the length of $\overline{SO}$ should be smaller than $\overline{ST}$.
% 변 ST는 삼각형 SOT의 가장 긴 변의 길이이므로, 변SO에 해당되는 entire voltage는 이 값보다 작은 값이어야 한다.
Thus, we can estimate the entire voltage of the graph as a positive value less than the sum of the minimum costs between adjacent levels.
% 따라서 우리는 그래프에 걸리는 전체 전압을 인접 레벨 사이의 최소 코스트의 합보다 작은 양의 값으로 추정할 수 있다.

\begin{df}\label{as2}
Let the ratio of the lengths between adjacent levels which divide $\overline{OT}$ be equal to the ratio of average costs between adjacent levels.
% 변 OT를 분할하는 레벨 구간별 길이의 비율은 레벨 구간별 평균 코스트들의 비율과 같다고 간주하자.
\end{df}
The $\overline{OT}$ is the sum of the lengths between adjacent levels:
% 변 OT는 다음과 같이 the lengths between adjacent levels들의 합으로 이루어져 있다.
\begin{equation}\label{eq9}
    \overline{OT}=\sum_{k=1}^N \nolimits L_{(k-1),(k)},
\end{equation}
where $L_{(k-1),(k)}$ is the lengths between level $k-1$ and level $k$.
% 여기서 L은 the lengths between level $k-1$ and level $k$의 값이다.
We do not estimate the length of $\overline{OT}$ directly as a specific value, but we estimate the ratio of the lengths between adjacent levels that divide $\overline{OT}$.
% 우리는 변 OT의 길이를 특정값으로 추정하진 않지만, 변 OT를 분할하는 the lengths between adjacent levels들 간의 비율을 추정한다.
The ratio of the lengths between adjacent levels is as follows:
% the lengths between adjacent levels 간의 비율은 다음과 같다:
\begin{equation}\label{eq10}
\begin{split}
    & L_{(0),(1)}:L_{(1),(2)}:\cdots:L_{(N-1),(N)}\\
    & =avg C_{(0),(1)}:avg C_{(1),(2)}:\cdots:avg C_{(N-1),(N)},
    \end{split}
\end{equation}
where $avg$ means the average value.
% 여기서 avg는 평균값을 의미한다.
According to the similarity in geometry, this ratio is equal to the ratio of the lengths between adjacent ideal points which divide $\overline{ST}$. 
% 도형의 닮음 규칙에 따라, 이 비율은 변 ST를 나누는 인접한 ideal point들 사이의 길이 비율과 같다.
Since we estimated the length of $\overline{ST}$ from~\eqref{eq8}, each appropriate length between adjacent ideal points also can be estimated with this ratio.
% 우리는 식8로 부터 ST의 길이를 추정했으므로, 각 the lengths between adjacent ideal points 사이의 적정 길이 역시 비율 정보로 부터 추정할 수 있다.
This can be interpreted as an appropriate small edge costs between adjacent levels.
% 이는 각 레벨 사이의 적정한 작은 코스트로 해석될 수 있다.

At the end of this section, we've shown the various transformations and interpretations of graphs through the concept of hypothetical electric potentials and also described hypothetical voltage estimation and appropriate edge cost estimation using two definitions.
% 이 절에서, 우리는 가상 전위의 개념을 통해 그래프의 다양한 변형과 해석을 보여 주었으며 또한 두 가지 가정을 사용하여 가상 전압 추정과 적절한 에지 비용 추정에 대해 설명했다.
The related example is in Sec.~\ref{cs}.
% 이에 대한 관련 예제는 case study 섹션에서 논한다.

\section{Graph Simplification using Kirchhoff's Circuit Law}
This section describes the graph simplification algorithm based on the graph analysis and the voltage estimation of the entire graph, that are described in the previous section.
% 이 섹션에서는, 앞 섹션에서 서술한 그래프 해석론과 그래프 전체의 전압 추정을 이용하여 그래프 간소화 알고리즘에 대해 설명한다.

\begin{algorithm}[t]
%\footnotesize
%\small
\caption{Graph Simplification}
\label{alg1}
\begin{algorithmic}[1]
\Statex $\hspace{-1.5em}\textbf{Volatage Estimation}$
\Statex $\hspace{-0.5em}\textbf{Input: } G(v,e)$ 
\Statex $\hspace{-0.5em}\textbf{Output: } V_{\max}$
\State Determine the level of each node in the given graph $G$.
\State Transform $G$ to level unit graph $G'(v',e)$.
\State Transform $G'$ to $3$-D potential graph and 
\Statex \hspace{1em} align in a straight line.
\State Construct a right triangle with three points: $O, S, T$.
\State Place ideal nodes of each level on $\overline{ST}$.
\State Estimate $\overline{ST}=\sum_{k=1}^N \min C_{(k-1),(k)}$.
\State Choose one real number $x\in\Pi$, where 
\Statex \hspace{1em} $\Pi=\{x|0<x<\overline{ST}\}$. % Pi에 속하는 실수 하나 선택.
\State \textbf{return} $x$.
\newline
\Statex $\hspace{-1.5em}\textbf{Edge Removal via Sub-current Estimation}$
\Statex $\hspace{-0.5em}\textbf{Input: } G(v,e), V_{\max}=x$ 
\Statex $\hspace{-0.5em}\textbf{Output: } G^{*}(v,e')$
\State Map $G$ to an electric circuit as $C(e_j)\rightarrow Resistance_j$.
\State Set the sub-currents of $Resistance_j$ as $I_j$.
\State Create the system of linear equations using 
\Statex \hspace{1em} KCL and KVL.
\State Solve the system of linear equations and obtain $I_1\cdots I_j$.
\For{int $i=1$; $i\leq j$; $i$++}{
    \If {$I_i=0$}
        \State Remove $e_i$ on $G$.
        \State $G=G_{removal,e_i}$.
    \EndIf
    }
\EndFor
\State \textbf{return} $G$.
\end{algorithmic}
\end{algorithm}

\subsection{Proposed Algorithm}
The algorithm consists of two processes, voltage estimation and edge removal via sub-current estimation.
% 이 알고리즘은 크게 전압 산정, 서브 전류 추정을 통한 엣지 제거, 두 가지 프로세스로 나뉜다.

A summary of the first process, voltage estimation, which described in the previous section as follows.
% 이전 섹션에서 서술한 첫번째 프로세스인 전압 산정 프로세스에 대해 요약하면 다음과 같다.
At first, determine the level of each node in the given graph $G(v,e)$, with $\abs{v}=i$ nodes and $\abs{e}=j$ edges, where $i$ and $j$ are positive integers.
% 먼저, 그래프의 노드별 레벨을 산정한다.
Construct a right triangle with three points: origin $O=(0,0,0)$, start node $S=(0,V_{max},0)$, and terminal node $T=(L,0,0)$, where $V_{max}$ and $L$ are positive real numbers, and $L<\overline{ST}$.
% 원점(0,0,0), 스타트 노드(0,V_max,0) , 터미널 노드(L,0,0)를 세 꼭지점으로 갖는 직각 삼각형을 구성한다, 여기서 L와 V_max는 양의 실수이며, L은 ST보다 작은값을 가진다.
Place ideal nodes of each level within $\overline{ST}$ in level order.
% 각 레벨의 이상적인 노드들을 레벨 순서대로 변 ST 내에 배치한다.
In other words, as shown in Fig.~\ref{fig5}, ideal nodes should be placed sequentially on the diagonal of the right triangle. 
% 다시 말하면, 그림 5에서와 같이 직각 삼각형의 대각선에 ideal node를 순차적으로 배치하라는 뜻이다.
Calculate the estimated length of $\overline{ST}$ as shown in~\eqref{eq8}.
% ST 거리를 식(8)과 같이 산정한다.
Choose one of any positive real number smaller than the length of $\overline{ST}$, and determined as $V_{\max}$.
% 산정된 원점-터미널노드 사이의 거리보다 작은 임의의 양의 실수를 V_max로 산정한다.
Normally, the value of $V_{\max}$ is chosen from positive integer values.
% 보통의 경우, V_max는 양의 정수 값 중에 고른다.

The second process, edge removal via sub-current estimation, is performed as follows.
% 두번째 프로세스인 edge removal via sub-current estimation의 과정은 다음과 같다.
Consider each edge costs (or weights) as each resistance, and consider $V_{max}$, which is determined in the first process, as an entire voltage of graph. 
% 엣지 코스트 (or weight)들을 저항으로 간주하고, voltage estimation process에서 결정한 V_max를 회로 전체의 저항으로 간주한다.
Set the sub-currents flowing at each edge of the graph as $I_1 \cdots I_j$, and create the system of linear equations using KCL and KVL.
% 그래프의 각 엣지에 흐르는 전류를 I_1 ~ I_j로 두고, KCL과 KVL을 이용하여 선형연립방정식을 만든다.
Solve the system of linear equations and find the values of the sub-currents $I_1 \cdots I_j$.
% 연립 방정식을 풀어, sub-currents $I_1 \cdots I_j$를 구한다.
To find the values of the sub-currents, the number of required linear equations is maximum $j$.
% 선형방정식은 최대 j가 필요할 것이다.
Remove the edges with zero sub-currents.
% 전류가 0인 구간의 엣지를 제거한다.

This algorithm is effective when there are many equipotential nodes with no current flowing between them.
% 해당 알고리즘은 사이에 전류가 흐르지 않는 등전위 노드가 많은 경우에 효과적이다.
In a complex graph, there is also a high probability of many equipotential nodes, so utilization is expected.
% 매우 복잡한 그래프에서는 등전위 노드가 많을 확률도 높으므로, 그 활용도가 가히 기대된다. 

\subsection{Case Study}\label{cs} 
In this section, we explain the graph simplification algorithm with a simple example.
% 이 섹션에서는 간단한 예제를 통해, Graph Simplification 알고리즘의 적용을 확인한다. 

\begin{figure}[t]
    \centering
    \includegraphics[width=.9\linewidth]{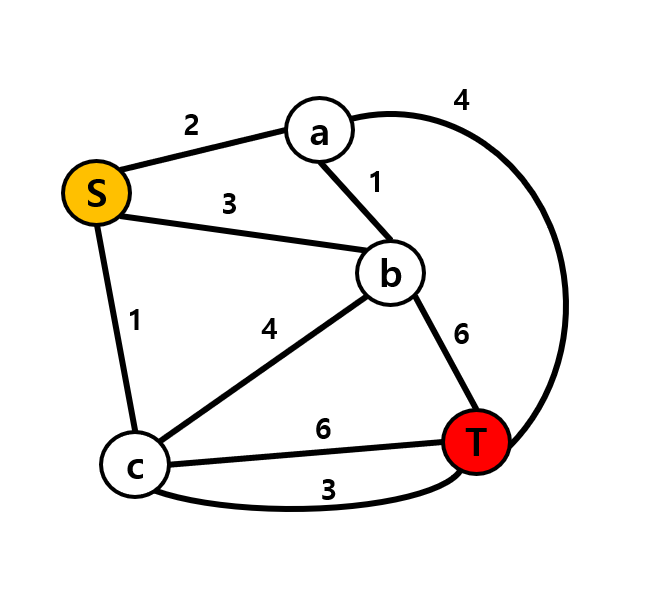}
    \caption{Weighted-graph $G(5,9)$. Each number next to each edge represents the edge cost (or weight).}
    \label{fig6}
\end{figure}

Suppose that, there is a weighted-graph, refer to Fig.~\ref{fig6}.
% 그림6과 같은 그래프가 있다고 가정하자.
We need to find the shortest path in a given graph $G(5,9)$, and we want to simplify this graph first.
% 우리는 이 그래프에서 최단 경로를 찾아야 하는데, 그래프를 단순화 시켜서 보고 싶다.
The levels of nodes $S$, $a$, $b$, $c$, and $T$ in this graph correspond to $0$, $1$, $1$, $1$, and $2$, respectively.
% 이 그래프의 노드 S, a, b, c, T의 레벨은 각각 0, 1, 1, 1, 2에 대응된다.
Thus, we can convert graph $G$ to level unit graph $G'(3,9)$.
% 따라서 우리는 그래프 G를 level unit graph G'(3,9)로 변환할 수 있다.
Through the process $3$ to $6$ of the algorithm~\ref{alg1}, a right triangle with points $S=(0,V_{\max},0)$, $O=(0,0,0)$, and $T=(L,0,0)$ can be formed, and the length of $\overline{ST}$ can be estimated as:
% 또한 알고리즘의 3~6의 과정을 통해, 직각삼각형 SOT를 형성할 수 있고, 변ST의 길이를 다음과 같이 추정할 수 있다.
\begin{equation}\label{eq11}
\begin{split}
    \overline{ST} & =\min C_{(0),(1)}+\min C_{(1),(2)}\\
    & =\min(2,3,1)+\min(4,6,6,3)=1+3=4. 
\end{split}
\end{equation}
Through the process $7$ to $8$ of Algorithm~\ref{alg1}, $V_{\max}$ can be a real number where $0<V_{\max}<4$, i.e., we set $V_{\max}=3$. 
% 알고리즘 7~8의 과정을 통해, V_max의 값은 0<x<4 사이에 있음을 알 수 있다. 

% 우리는 V_max의 값으로 3을 선택한다.
\begin{figure}[t] 
    \centering
    \includegraphics[width=.9\linewidth]{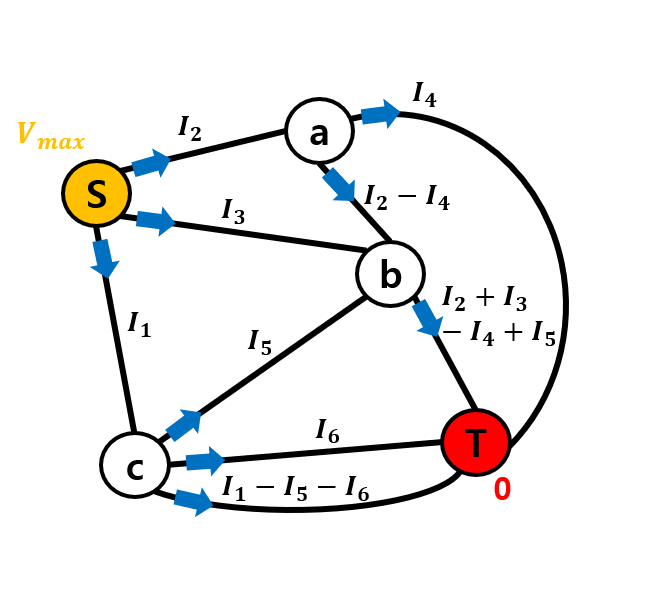}
    \caption{Graph of sub-current applying KCL. The current flows from $S$ with potential $V_{max}$ to $T$ with potential zero.}
    % KCL을 적용하여 sub-current를 나타낸 그래프. 시작노드 S는 V_max의 전위를 갖고, 터미널 노드 T는 0의 전위를 갖는다.
    \label{fig7}
\end{figure}

\begin{figure}[t]
    \centering
    \includegraphics[width=.9\linewidth]{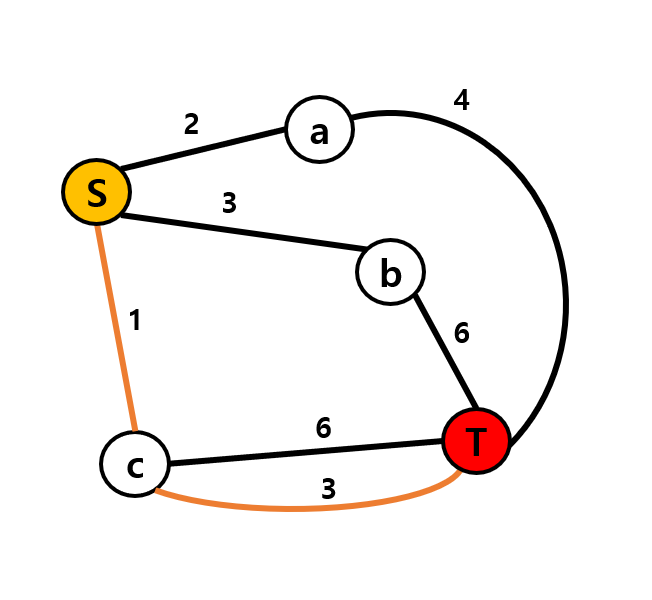}
    \caption{Graph $G^{*}(5,7)$, which is the output of the graph simplification algorithm. The orange path is the shortest path (or the minimum cost path).}
    % 그래프 간소화 알고리즘의 아웃풋인, 간소화된 그래프 G^{*}
    % 주황색 경로는 최단경로 (최소 비용 경로)를 나타낸다.
    \label{fig8}
\end{figure}

We consider each edge cost as each resistance and find the sub-current flowing through each edge.
% 우리는 이제 엣지코스트를 저항으로 간주하여, 각 엣지에 흐르는 전류를 구한다.
Fig.~\ref{fig7} shows the sub-currents flowing through each edge by applying KCL.
% KCL을 적용하여, 각 엣지에 흐르는 서브 전류를 나타내면, 그림7과 같다.
Since there are $6$ sub-currents in this graph, we may create the system of linear equations with $6$ closed-circuit loops according to KVL are as follows:
% sub-current는 총 6개이므로, KVL에 따라 폐회로 6개에서 다음과 같이 연립방정식을 세우면 다음과 같다.
\begin{equation}\label{eq12}
\begin{split}
    \begin{cases} 
    & 3=2I_2+4I_4,\\
    & 3=I_1+6I_6,\\
    & 3=I_1+3(I_1-I_5-I_6),\\
    & 3=3I_3+6(I_2+I_3-I_4+I_5),\\
    & 0=3I_3-4I_5-I_1,\\
    & 0=2I_2+(I_2-I_4)-3I_3.
    \end{cases}
\end{split}
\end{equation}
The values of sub-currents $I_1$, $I_2$, $I_3$, $I_4$, $I_5$, and $I_6$ are obtained from~\eqref{eq12} as $1$, $\frac{1}{2}$, $\frac{1}{3}$, $\frac{1}{2}$, $0$, and $\frac{1}{3}$, respectively.
% 이 연립방정식을 풀면, 각 전류의 값은 다음과 같이 구해진다 ~~~.
%%%%%%%%%%%%%%%%%%%%% 제거 수식 %%%%%%%%%%%%%%%%%%%%%
%\begin{equation}\label{eq13}
%\begin{split}
%    & I_{sa}=I_2=\frac{1}{2}\\
%    & I_{sb}=I_3=\frac{1}{3}\\
%    & I_{sc}=I_1=1\\
%    & I_{ab}=I_2-I_4=0\\
%    & I_{cb}=I_5=0\\
%    & I_{aT}=I_4=\frac{1}{2}\\
%    & I_{bT}=I_I_2+I_3-I_4+I_5=\frac{1}{3}\\
%    & I_{cT}=I_6=\frac{1}{3}\\
%    & I_{cT}=I_1-I_5-I_6=\frac{2}{3}
%\end{split}
%\end{equation}
%%%%%%%%%%%%%%%%%%%%% 제거 수식 %%%%%%%%%%%%%%%%%%%%%
We confirm that there is no sub-current flowing through the edges $ab$ and $cb$ (i.e., $I_{ab}=I_2-I_4=0$ and $I_{cb}=I_5=0$), so they will be removed.
% 우리는 이를 통해 엣지 ab와 bc에 전류가 흐르지 않음을 알 수 있고, 제거 가능하다.
Fig~\ref{fig8} shows a simplified graph $G^{*}(5,7)$ with two edges removal.
% 두 엣지를 잘라내고 간소화된 그래프는 그림 8과 같다.
In $G^{*}$, there are only four paths from $S$ to $T$, i.e., we can easily find the shortest path.
% 간소화 된 그래프에는 S-T 까지의 경로가 단순한 네 가지 뿐이고, 우리는 원본 그래프 G에서 보다 쉽게 최단 경로를 찾을 수 있다.

In addition, due to \textit{Definition}~\ref{as2}, the ratio of the lengths between adjacent level which divides $\overline{OT}$ is obtained as:
% 추가로 가정2에 따라 이 예시에서 the ratio of the lengths between adjacent level which divide OT를 구하면 다음과 같다.
\begin{equation}\label{eq13}
\begin{split}
    L_{(0),(1)}:L_{(1),(2)} & =avg C_{(0),(1)}:avg C_{(1),(2)}\\
    & =avg(2,3,1):avg(4,6,6,3)\\
    & = 2:\frac{19}{4}=8:19.
    \end{split}
\end{equation}
This ratio is equal to the ratio of the lengths between adjacent ideal points which divide $\overline{ST}$, so $\overline{SD}$ and $\overline{DT}$ are as:
% 이 비율은 the ratio of the lengths between adjacent ideal points which divide ST와 같으므로, ST를 분할하는 SD, DT는 다음과 같다.
\begin{equation}\label{eq14}
\begin{split}
    & \overline{SD}=\left(\frac{8}{8+19}\right)\overline{ST}\approx 1.185, \\
    & \overline{DT}=\left(\frac{19}{8+19}\right)\overline{ST}\approx 2.815,
\end{split}
\end{equation}
where $D$ is an ideal point of level $1$ and $\overline{ST}=4$ is an obtained value from~\eqref{eq11}.
% 여기서 D는 레벨1의 이상적인 지점이고, ST=4로 식11에서 구한 값이다.
$\overline{SD}$ and $\overline{DT}$ can be consider as relatively small edge costs between level $0$ to $1$ and level $1$ to $2$, respectively.
%  $\overline{SD}$ and $\overline{DT}$는 각각 레벨0-1 사이의 적정 최단 경로 추정치, 레벨1-2 사이의 적정 최단 경로 추정치로 볼 수 있는데, 
In Fig.~\ref{fig6}, the closest to $\overline{SD}=1.185$ among the edge costs between level $0$ to $1$ is $1$ and the closest to $\overline{DT}=2.815$ among the edge costs between levels $1$ to $2$ is $3$.
% 실제로 레벨 0-1 사이의 엣지 코스트 2,3,1 중 가장 SD에 가장 근사한 값은 1이고 레벨 1-2 사이의 엣지 코스트 4,6,6,3 중 가장 DT에 근사한 값은 3이며, 
These two edges correspond to the actual shortest path, represented by the orange path in Fig.~\ref{fig8}.
% 이 두 값은 그림8에서 주황색 경로로 표현되는 실제 최단 경로와 일치한다.
This is because, in this example, all the edges between the same level nodes are removed by the graph simplification algorithm.
% 이는 해당 예제에서 간소화 알고리즘 과정을 통해 동레벨 노드간의 경로가 모두 제거되었기 때문이다.

\section{Conclusions and Future Work} 
In this paper, we have proposed graph analysis methods using electric potentials, such as level unit transformed graph analysis, straight-line alignment $3$-D potential graph analysis, and geometric analysis.
Based on these analysis methods, a graph simplification algorithm using Kirchhoff's Circuit Law has been proposed.
The graph simplification algorithm consists of two processes, i.e., voltage estimation and edge removal via sub-current estimation.
We have applied this algorithm to the example graph, and have demonstrated its usefulness as a pre-processing algorithm for path-finding problems.
Thus, we also have confirmed the validity of the definitions underlying the proposed graph simplification algorithm.

As a future work direction, we will figure out which kinds of applications can be useful; and then conduct data-intensive performance evaluations.

\section*{Acknowledgment}
This research was supported by National Research Foundation of Korea (2019M3E4A1080391). % 양자과제
J. Kim is a corresponding author (e-mail: joongheon@korea.ac.kr).

\bibliographystyle{IEEEtran}
\bibliography{main}

\IEEEtriggeratref{3}
%%
%% which triggers a \newpage (i.e., new column) just before the given
%% reference number. Note that you need to adapt this if you modify
%% the paper.  The "triggered" command can be changed if desired:
%%
%\IEEEtriggercmd{\enlargethispage{-20cm}}
%%
%%%%%%

%%%%%%
%% References:
%% We recommend the usage of BibTeX:
%%
%\bibliographystyle{IEEEtran}
%\bibliography{definitions,bibliofile}
%%
%% where we here have assume the existence of the files
%% definitions.bib and bibliofile.bib.
%% BibTeX documentation can be obtained at:
%% http://www.ctan.org/tex-archive/biblio/bibtex/contrib/doc/
%%%%%%

%% Or you use manual references (pay attention to consistency and the
%% formatting style!):

\end{document}